# $H_{c2}$ of anisotropy two-band superconductors by Ginzburg-Landau approach

P.Udomsamuthirun[1], A.Changjan [1],
C.Kumvongsa [2] and S.Yoksan[1]

[1] Department of Physics, Faculty of Science, Srinakharinwirot University, Bangkok 10110, Thailand. E-mail: udomsamut55@yahoo.com

[2] Department of Basic Science, School of Science, The University of the Thai Chamber of Commerce, Dindaeng, Bangkok 10400, Thailand.

## Abstract

The purpose of this research is to study the upper critical field ($H_{c2}$) of two-band superconductors by two-band Ginzburg - Landau approach. The analytical formula of $H_{c2}$ included anisotropy of order parameter and anisotropy of effective-mass are found. The parameters of the upper critical field in ab-plane($H_{c2}^{//ab}$) and c-axis($H_{c2}^{//c}$) can be found by fitting to the experimental data. Finally, we can find the ratio of upper critical field that temperature dependent in the range of experimental result.

# 1. Introduction

The muti-band characteristic of the superconducting state in MgB$_2$ [1-3] and nonmagnetic borocarbides(LuNi$_2$B$_2$C and Y Ni$_2$B$_2$C) [4] and NbSe$_2$[5] are clearly evident in many measurement. In contrast to conventional superconductors, the upper critical field for a bulk MgB$_2$, LuNi$_2$B$_2$C and Y Ni$_2$B$_2$C has a positive curvature near T$_c$[6-8]. Canfield and Crabtree[9] show the anisotropy of upper critical field of MgB$_2$. The anisotropy, quantified as the ratio $\gamma = \dfrac{H_{c2}^{//ab}}{H_{c2}^{//c}}$, is not only large, but it has an unusual temperature dependence. Askerzade [10] study H$_{c1}$(T) and H$_{c2}$(T) for MgB$_2$ and nonmagnetic borocarbides by using the isotropy two-band Ginzberg-Landau model. He shows the complicate formula of critical magnetic field. In the H$_{c2}$ consideration, his formula is less fit to experimental data than Drechsler 's[11]. In the recent work of Dao and Zhitomirsky[12], they study the effect of angular and temperature on upper critical field of MgB$_2$ determined with the anisotropy two-band Ginzberg-Landau theory. They found that the temperature variation of the ratio of two gaps is responsible for the upward temperature dependence of in-plane H$_{c2}$ as well as for the deviation of its out-of-plane behavior from the standard angular dependence.

In this paper, we study the upper critical magnetic field($H_{c2}$) of anisotropy two-band s-wave superconductors by Ginzburg-Landau theory. We can find the formula of H$_{c2}$ included anisotropy of order parameter and anisotropy of effective-mass. The upper critical field in ab-plane($H_{c2}^{//ab}$) and c-axis($H_{c2}^{//c}$) can be found by fitting to the experimental data of MgB$_2$. The anisotropy parameter $\gamma = \dfrac{H_{c2}^{//ab}}{H_{c2}^{//c}}$ are shown finally.

# 2. Model and calculation

In the presence of two order parameters, $\psi_1$ and $\psi_2$, in a superconductor, the Ginzburg-Landau free energy can be written as[10,13-15]

$$F[\psi_1,\psi_2] = \int d^3r (F_1(\psi_1) + F_2(\psi_2) + F_{12}(\psi_1,\psi_2) + \frac{H^2}{8\pi}) \qquad (1)$$

with

$$F_j(\psi_j) = \frac{\hbar^2}{4m_j}\left|(\nabla - \frac{2\pi i A}{\phi_0})\psi_j\right|^2 + \alpha_j(T)\psi_j^2 + \frac{\beta_j}{2}\psi_j^4 \qquad (2)$$

$$F_{12}(\psi_1,\psi_2) = \varepsilon(\psi_1\psi_2^* + c.c.) + \varepsilon_1[(\nabla + \frac{2\pi i A}{\phi_0})\psi_1^*(\nabla - \frac{2\pi i A}{\phi_0})\psi_2 + c.c.] \qquad (3)$$

Fj is the free energy of separate bands. $\vec{H}$ is the external magnetic field($\vec{H} = \nabla x \vec{A}$). $\alpha_j, \beta_j$, m$_j$ are the temperature-dependent coefficient, temperature-independent coefficient and mass of the carriers in band j (j=1,2) respectively. $\alpha_j = \lambda_j(T - T_c)$ where $\lambda_j$ is the temperature-independent constant. The coefficient $\varepsilon, \varepsilon_1$ are the interband mixing of two order parameters and their gradients that $\varepsilon_1$ is equal to zero in Dao and Zhitomirsky[12] s'free energy.

Let there has only magnetic field in z direction, $\vec{H} = H\hat{k}$, and vector potential is in y direction, $\vec{A} = Hx\hat{j}$ that $H = |\vec{H}|$ and x is the displacement. Minimization of the free energy Eq. (1) with respect to the order parameters, $\psi_1^*$ and $\psi_2^*$, we get[10,13]

$$(\alpha_1(T) + \frac{\hbar He}{2m_1 c})(\alpha_2(T) + \frac{\hbar He}{2m_2 c}) = (\varepsilon - \frac{2He\varepsilon_1}{\hbar c})^2 \qquad (4)$$

Let $m_1 = m_2 = m$, we get

$$(\frac{\hbar^2 e^2}{4m^2 c^2} - \frac{4\varepsilon_1^2 e^2}{\hbar^2 c^2})H_{c2}^2 + ((\alpha_1 + \alpha_2)\frac{\hbar e}{2mc} + \frac{4\varepsilon\varepsilon_1 e}{\hbar c})H_{c2} + \alpha_1\alpha_2 - \varepsilon^2 = 0 \qquad (5)$$

We set new parameters to scale Eq.(5), $\xi_1 = \sqrt{\frac{\hbar^2}{2m\alpha_1}}$, $\xi_2 = \sqrt{\frac{\hbar^2}{2m\alpha_2}}$, $\xi_{12} = \sqrt{\frac{\hbar^2}{2m\varepsilon}}$ that are the 1$^{st}$ band, 2$^{nd}$ band and interband effective coherence length, respectively. $\varepsilon_1 = \frac{\kappa\hbar^2}{4m}$ where $\kappa$ is the gradients of interband mixing of two order parameters in energy unit and $\phi_0 = \frac{2\pi\hbar c}{e}$ that is quantum flux. We find

$$\frac{\pi^2}{\phi_0^2}(1-\kappa^2)H_{c2}^2 + \frac{\pi}{\phi_0}\left(\frac{1}{2\xi_1^2} + \frac{1}{2\xi_2^2} + \frac{\kappa}{\xi_{12}^2}\right)H_{c2} + \frac{1}{4}\left(\frac{1}{\xi_1^2 \cdot \xi_2^2} - \frac{1}{\xi_{12}^4}\right) = 0 \qquad (6)$$

By making the approximation within the condition, $0.5 < \kappa < 1$, and keeping some of the lower order term, the analytical equation of upper critical field is

$$H_{c2} = -\frac{\phi_0}{2\pi\xi_{12}^2}(\frac{\xi_{12}^2}{\xi_1^2} + \frac{\xi_{12}^2}{\xi_2^2} + 2\kappa)[\frac{1}{1-\kappa^2} - \frac{(\frac{\xi_{12}^2}{\xi_1^2\xi_2^2}-1)}{(\frac{\xi_{12}^2}{\xi_1^2} + \frac{\xi_{12}^2}{\xi_2^2} + 2\kappa)^2} + \frac{(1-\kappa^2)(\frac{\xi_{12}^4}{\xi_1^2\xi_2^2}-1)^2}{(\frac{\xi_{12}^2}{\xi_1^2} + \frac{\xi_{12}^2}{\xi_2^2} + 2\kappa)^4}] \qquad (7)$$

The derivation of Eq.(7) from Eq.(6) is shown in an appendix.

Eq.(7) can be reduced to $H_{c2}$ of single band superconductors, $H_{c2} = \frac{\phi_0}{2\pi\xi^2}$, by taking $\alpha_2 = \varepsilon = \varepsilon_1 = 0$.

**3. Anisotropic mass tensor model**

Let us consider the two-band s-wave superconductor that shows the layered property. In the layered superconductors, since the overlap of electron wave function is larger within the layers than between layers, it can be assumed that the electrons have a high effective mass for motion normal to the layers and a low effective mass for motion within a layer. Let the mass tensor $\{m\}$ has the form [16]

$$\{m\} = \begin{pmatrix} m & 0 & 0 \\ 0 & m & 0 \\ 0 & 0 & M \end{pmatrix} \qquad (8)$$

where $M > m$. Since the coherence length $\xi$ depends on the effective mass as $\xi \alpha \frac{1}{\sqrt{m}}$, the coherence length also becomes a tensor and given by

$$\{\xi\} = \begin{pmatrix} \xi & 0 & 0 \\ 0 & \xi & 0 \\ 0 & 0 & \delta\xi \end{pmatrix} \quad (9)$$

where $\delta = \sqrt{m/M}$.

The effect of anisotropy mass tensor on the upper critical field is included by replace $\xi$ in Eq.(7) with coherence length tensor $\{\xi\}$. The calculation procedure is as same as reference [16], then we get

$$H_{c2} = -\frac{\phi_0}{2\pi\xi_{12}^2(\sin^2\theta + \delta^2\cos^2\theta)}(\frac{\xi_{12}^2}{\xi_1^2}+\frac{\xi_{12}^2}{\xi_2^2}+2\kappa)[\frac{1}{1-\kappa^2} - \frac{(\frac{\xi_{12}^2}{\xi_1^2\xi_2^2}-1)}{(\frac{\xi_{12}^2}{\xi_1^2}+\frac{\xi_{12}^2}{\xi_2^2}+2\kappa)^2}$$

$$+\frac{(1-\kappa^2)(\frac{\xi_{12}^4}{\xi_1^2\xi_2^2}-1)^2}{(\frac{\xi_{12}^2}{\xi_1^2}+\frac{\xi_{12}^2}{\xi_2^2}+2\kappa)^4}] \quad (10)$$

Here $\theta$ is the angle between magnetic field and the layer of superconductor.

We consider the MgB$_2$ superconductor. MgB$_2$ has a hexagonal crystal structure and the anisotropy is most likely to occur along the c-direction similar to other hexagonal crystal such as U Pt$_3$ and U Pd$_2$Al$_3$. We can concluded that MgB$_2$ is a layered superconductor with anisotropy by the nature of the crystal structure. It is seem that H$_{C2}$ of MgB$_2$ can be described by Eq.(10). On the recent paper of Koshelev and Golubov[17], they propose that MgB$_2$ can not be described by the anisotropy mass Ginzberg-Landau theory. The anisotropy mass Ginzberg-Landau theory has a temperature-limited to describe H$_{c2}$ of MgB$_2$ less than 2% away from T$_c$. Since such a small temperature region is difficult to accurately probe experimentally, one may say that there is a complete breakdown of anisotropic mass Ginzberg-Landau theory in MgB$_2$. Although this model can not describe the H$_{c2}$ of MgB$_2$ superconductor, we think that this model should be suitable for the other two-band s-wave superconductors.

## 4. Anisotropic order parameters

Here we make the assumption that order parameters are proportioned to the energy gaps, $\psi \; \alpha \; \Delta$, [18]. We can write the order parameters in the form

$$\psi = \psi_0(T)f(\vec{k}) \quad (11)$$

where $f(\vec{k})$ is the anisotropy function. $\vec{k}$ is the wave vector. In MgB$_2$ superconductor, there are many kinds of anisotropy function proposed by researchers [19-21].

Substitution Eq.(11) into Eq.(1), we can get the formula of H$_{c2}$ as

$$H_{c2} = -\frac{\phi_0}{2\pi\xi_{12}'^2}(\frac{\xi_{12}'^2}{\xi_1^2}+\frac{\xi_{12}'^2}{\xi_2^2}+2\sqrt{\Omega}\,\kappa)[\frac{1}{1-\Omega\kappa^2} - \frac{(\frac{\xi_{12}'^2}{\xi_1^2\xi_2^2}-1)}{(\frac{\xi_{12}'^2}{\xi_1^2}+\frac{\xi_{12}'^2}{\xi_2^2}+2\sqrt{\Omega}\,\kappa)^2}$$

$$+ \frac{(1-\Omega\kappa^2)(\frac{\xi_{12}'^4}{\xi_1^2\xi_2^2}-1)^2}{(\frac{\xi_{12}'^2}{\xi_1^2}+\frac{\xi_{12}'^2}{\xi_2^2}+2\sqrt{\Omega}\,\kappa)^4}] \quad (12)$$

where $\Omega = \frac{\langle f_1(\hat{k})f_2(\hat{k})\rangle^2}{\langle f_1^2(\hat{k})\rangle\langle f_2^2(\hat{k})\rangle}$ and $\xi_{12}' = \sqrt{\frac{\hbar^2}{2m\varepsilon_1\sqrt{\Omega}}} = \frac{\xi}{\Omega^{1/4}}$.

$<.....>$ is the averaged over Fermi surface.

$f_1(k)$ and $f_2(k)$ are the anisotropy function of first and second band.

In case of the symmetry gap $f_1(k) = f_2(k) = 1$, that gives $\Omega = 1$. Eq.(12) will be reduced to be Eq.(7). Let us consider the anisotropy model of Haas and Maki[19], $f(\theta) = \frac{1+a'\cos^2\theta}{1+a'}$. Here $\theta$ is the polar angle and $a'$ is anisotropy parameter. We assume that both band have the same anisotropy function, but they have the difference in anisotropy parameter. $f_1(\theta) = \frac{1+a'\cos^2\theta}{1+a'}$ and $f_2(\theta) = \frac{1+b'\cos^2\theta}{1+b'}$. In this model, the averaged over Fermi surface is

$<f(\theta)> = \frac{1}{2}\int_0^\pi d\theta\sin\theta\, f(\theta)$ that $<f_1^2(\theta)> = \frac{15+10a'+3a'^2}{15(1+a')^2}$ and

$<f_1(\theta)f_2(\theta)> = \frac{5(3+b')+a'(5+3b')}{15(1+a')(1+b')}$. We get $\Omega = \frac{(5(3+b')+a'(5+3b'))^2}{(15+a'(10+3a'))(15+b'(10+3b'))}$. If $a' = b'$, we get $\Omega = 1$. This means that if each band has the same anisotropy function and anisotropy parameter, the anisotropy will show no effect on $H_{c2}$. For our consideration, $\Omega$ is dependent on anisotropy parameters. If we know the values of anisotropy parameters, we can get the value of $\Omega$. For simplify, we will consider $\Omega$ as a constant parameter.

We consider the experimental data of MgB$_2$ and make the assumption that the upper critical field in ab-plane ($H_{c2}^{//ab}$) and c-axis ($H_{c2}^{//c}$) can be found by set of the suitable parameters. We can write Eq.(12) in form that $H_{c2} = H_{c2}(a_1, a_2, \Omega, \kappa)$, where $a_1 = \frac{\hbar^2}{2m\lambda_1}$, $a_2 = \frac{\hbar^2}{2m\lambda_2}$. After fitting Eq.(12) to the experimental data of upper critical field of single crystal MgB$_2$[22-24], the upper critical field in ab-plane ($H_{c2}^{//ab}$) is $H_{c2}^{//ab} = H_{c2}(6.5, 6.5, 0.7, 0.5)$, and in c-axis ($H_{c2}^{//c}$) is $H_{c2}^{//c} = H_{c2}(13, 13, 0.7, 0.5)$, as shown in Fig.(1). In Figure.(2), the ratio of upper critical field (

$\gamma = \frac{H_{c2}^{//ab}}{H_{c2}^{//c}} = \frac{H_{c2}(6.5, 6.5, 0.7, 0.5)}{H_{c2}(13, 13, 0.7, 0.5)}$) versus temperature is shown. The maximum ratio $\gamma$ at T=0 K is equal to 2.0. Our result is in range of 1 to 13 [20] and when temperature is increased, $\gamma$ is decreased. (although, it is almost constant in low temperature region). This behavior agree with the result of Miranovic′, Machida and Kogan[21] and Canfield and Crabtree[9] . .

**5. Conclusion**

The effect of an anisotropy in mass tensor and an anisotropy of order parameters on upper critical field are considered in our model. We use model of an anisotropic order parameters on upper critical field in our numerical calculation to compare to experimental data of $MgB_2$. The parameters of the upper critical field in ab-plane( $H_{c2}^{//ab}$ ) and c-axis( $H_{c2}^{//c}$ ) can be found by fitting to the experimental data. Finally, we can find the ratio of upper critical field that depend on temperature in the range of experimental result. Our model can find the ratio of $\gamma = \dfrac{H_{c2}^{//ab}}{H_{c2}^{//c}}$ at T=0 K equal to 2.0 .

**Acknowledgement** The author would like to thank Thailand Research Fund for financial support, and the university of the Thai Chamber of Commerce and Srinakhariwirot university for partial financial support .

**Appendix**

In this appendix, we would like to show how to get $H_{c2}$ equation(Eq.(7)) from Eq.(6). And this procedure can be apply to get solution as Eq.(10) and (12) also. Consider Eq.(6) that

$$\frac{\pi^2}{\phi_0^2}(1-\kappa^2)H_{c2}^2 + \frac{\pi}{\phi_0}\left(\frac{1}{2\xi_1^2}+\frac{1}{2\xi_2^2}+\frac{\kappa}{\xi_{12}^2}\right)H_{c2} + \frac{1}{4}\left(\frac{1}{\xi_1^2 \cdot \xi_2^2}-\frac{1}{\xi_{12}^4}\right) = 0 \tag{6}$$

Since Eg.(6) is a quadratic equation, we can get solution as

$$H_{c2} = -\frac{\phi_0}{2\pi(1-\kappa^2)}\left(\frac{1}{2\xi_1^2}+\frac{1}{2\xi_2^2}+\frac{\kappa}{\xi_{12}^2}\right) \pm \frac{\phi_0}{2\pi(1-\kappa^2)}\left(\left(\frac{1}{2\xi_1^2}+\frac{1}{2\xi_2^2}+\frac{\kappa}{\xi_{12}^2}\right)^2 - (1-\kappa^2)\left(\frac{1}{\xi_1^2\xi_2^2}-\frac{1}{\xi_{12}^4}\right)\right)^{1/2} \tag{A1}$$

Eq.(A1) has two terms in square root that we can make the approximation by considering these two terms. It can be considered into two cases.

<u>Case I</u>   For $(\frac{1}{2\xi_1^2}+\frac{1}{2\xi_2^2}+\frac{\kappa}{\xi_{12}^2})^2 \gg (1-\kappa^2)(\frac{1}{\xi_1^2\xi_2^2}-\frac{1}{\xi_{12}^4})$.

By making the approximation, Eq.(A1) become

$$H_{c2}^{(-)} = -\frac{\phi_0}{2\pi(1-\kappa^2)}\left(\frac{1}{2\xi_1^2}+\frac{1}{2\xi_2^2}+\frac{\kappa}{\xi_{12}^2}\right) + \left(\frac{1}{2}\right)\frac{\phi_0}{2\pi}\frac{(\frac{1}{\xi_1^2\xi_2^2}-\frac{1}{\xi_{12}^4})}{(\frac{1}{2\xi_1^2}+\frac{1}{2\xi_2^2}+\frac{\kappa}{\xi_{12}^2})} - \left(\frac{1}{8}\right)\frac{\phi_0}{2\pi}\frac{(1-\kappa^2)(\frac{1}{\xi_1^2\xi_2^2}-\frac{1}{\xi_{12}^4})^2}{(\frac{1}{2\xi_1^2}+\frac{1}{2\xi_2^2}+\frac{\kappa}{\xi_{12}^2})^3}$$

$$\tag{A2}$$

and

$$H_{c2}^{(+)} = -(\frac{1}{2})\frac{\phi_0}{2\pi}\frac{(\frac{1}{\xi_1^2\xi_2^2}-\frac{1}{\xi_{12}^4})}{(\frac{1}{2\xi_1^2}+\frac{1}{2\xi_2^2}+\frac{\kappa}{\xi_{12}^2})} + (\frac{1}{8})\frac{\phi_0}{2\pi}\frac{(1-\kappa^2)(\frac{1}{\xi_1^2\xi_2^2}-\frac{1}{\xi_{12}^4})^2}{(\frac{1}{2\xi_1^2}+\frac{1}{2\xi_2^2}+\frac{\kappa}{\xi_{12}^2})^3} \quad (A3)$$

<u>Case II</u>  For $(\frac{1}{2\xi_1^2}+\frac{1}{2\xi_2^2}+\frac{\kappa}{\xi_{12}^2})^2 << (1-\kappa^2)(\frac{1}{\xi_1^2\xi_2^2}-\frac{1}{\xi_{12}^4})$.

By making the approximation, Eq.(A1) become

$$H_{c2}^{(-)} = -\frac{\phi_0}{2\pi(1-\kappa^2)}(\frac{1}{2\xi_1^2}+\frac{1}{2\xi_2^2}+\frac{\kappa}{\xi_{12}^2}) - \frac{\phi_0}{2\pi}\sqrt{\frac{\frac{1}{\xi_{12}^4}-\frac{1}{\xi_1^2\xi_2^2}}{1-\kappa^2}} - \frac{\phi_0}{4\pi}\frac{(\frac{1}{2\xi_1^2}+\frac{1}{2\xi_2^2}+\frac{\kappa}{\xi_{12}^2})^3}{(1-\kappa^2)^{3/2}\sqrt{\frac{1}{\xi_{12}^4}-\frac{1}{\xi_1^2\xi_2^2}}}$$

(A4)

and

$$H_{c2}^{(+)} = \frac{\phi_0}{2\pi(1-\kappa^2)}(\frac{1}{2\xi_1^2}+\frac{1}{2\xi_2^2}+\frac{\kappa}{\xi_{12}^2}) + \frac{\phi_0}{2\pi}\sqrt{\frac{\frac{1}{\xi_{12}^4}-\frac{1}{\xi_1^2\xi_2^2}}{1-\kappa^2}} + \frac{\phi_0}{4\pi}\frac{(\frac{1}{2\xi_1^2}+\frac{1}{2\xi_2^2}+\frac{\kappa}{\xi_{12}^2})^2}{(1-\kappa^2)^{3/2}\sqrt{\frac{1}{\xi_{12}^4}-\frac{1}{\xi_1^2\xi_2^2}}}$$

(A5)

We have obtained four formula for the solution of Eq.(6). Next step, we will check the conditions to get physical solution of Eq.(6). We have the critical magnetic field of one band model as $H_{c2} = \frac{\phi_0}{2\pi\xi^2}$. Two band model can be reduced to one band model if we set $\alpha_2 = \varepsilon = \varepsilon_1 = 0$. Eq. (A3) is not reduced to the one-band's result. So Eq(A2),Eq.(A4),and Eq.(A5) can be the solution of two band model. We assume the next condition that $\frac{1}{2} < \kappa < 1$. Only Eq.(A2) satisfies this condition. Finally, the analytical equation of upper critical field is Eq.(A2) and we can rewrite to be

$$H_{c2} = -\frac{\phi_0}{2\pi\xi_{12}^2}(\frac{\xi_{12}^2}{\xi_1^2}+\frac{\xi_{12}^2}{\xi_2^2}+2\kappa)[\frac{1}{1-\kappa^2} - \frac{(\frac{\xi_{12}^2}{\xi_1^2\xi_2^2}-1)}{(\frac{\xi_{12}^2}{\xi_1^2}+\frac{\xi_{12}^2}{\xi_2^2}+2\kappa)^2} + \frac{(1-\kappa^2)(\frac{\xi_{12}^4}{\xi_1^2\xi_2^2}-1)^2}{(\frac{\xi_{12}^2}{\xi_1^2}+\frac{\xi_{12}^2}{\xi_2^2}+2\kappa)^4}] \quad (7)$$

**Reference**

1. X.K.Chen,M.J.Konstantinovich,J.C.Irwin,D.D.Lawrie,J.P.Frank,Phys.Rev.Lett. **87**,157002 (2001).
2. H.Giublio et al.,Phys.Rev.Lett. **87**,177008(2001).
3. F.Boquet et al., Phys.Rev.Lett. **87**,047001(2001).
4. M.Heinecke,K.Winzer,Z.Phys. B **98**,147(1995).
5. T.Yokoya et al.,Science **294**,2518(2001)
6. J.Freudenberger et al., Physica C **306**,1(1998).
7. K.H.Muller et al., J.Alloys Compd. **322**,L10(2001).
8. S.Bud'ko et al., Phys.Rev. B 63,220503(R)(2001).
9. P.C.Canfield and G.W.Crabtree,Physics Today,March 2003,34-40.
10. I.N.Askerzade,Physica C **397**,99(2003).
11. S.L.Drechsler, H.Rosner,S.V.Shulga, H.Eschrig, 2001,"High-$T_c$ Superconductors and Related materials",Ed. S.L.Drechsler and T.Mishonou(Kluwer Academic Publishers).
12. V.H.Dao and M.E.Zhitomirsky,European Physical Journal B **44**,183(2005).
13  I.N.Askerzade, A.Gencer,N.Guclu, Supercond.Sci.Technol. **15** ,L13-L16 (2002) .
14. I.N.Askerzade, A.Gencer,N.Guclu, A.Kihc, Supercond.Sci.Technol. **15**, L17 (2002)
15. I.N.Askerzade,A.Gencer,J.Phys.Soc.Jpn. **71**,1637(2002) .
16. R.C.Morris ,R.V.Coleman, and R.Bhandari,Phys.Rev.B **5**,895(1972).
17. A.E.Koshelev and A.A.Golubov,Physical Review Letter **92**,107008(2004).
18. A.L. Fetter; and J.D. Walecka.(1995). "Quantum theory of Many – Particle system" , International edition, MaGraw - Hill .
19. S.Haas and K. Maki, Phys.Rev.B **65**,020502(R)(2002).
20. V.G. Kogan and S.L.Bud'ko, Physica C **385** : 131(2003)
21. P.Miranovic, K. Machida, and V.G.Kogan, J. Phys.Soc.Jpn. **72** , 221(2003)
22. M.Xu et.al.,Appl.Phys.Lett. **79**,2779(2001).
23. S.Lee  et al.,J.Phys.Soc.Jpn. **70**,2255(2001).
24. O.F.de Lima et al.,Phys.Rev.Lett. **86**,5974(2001).

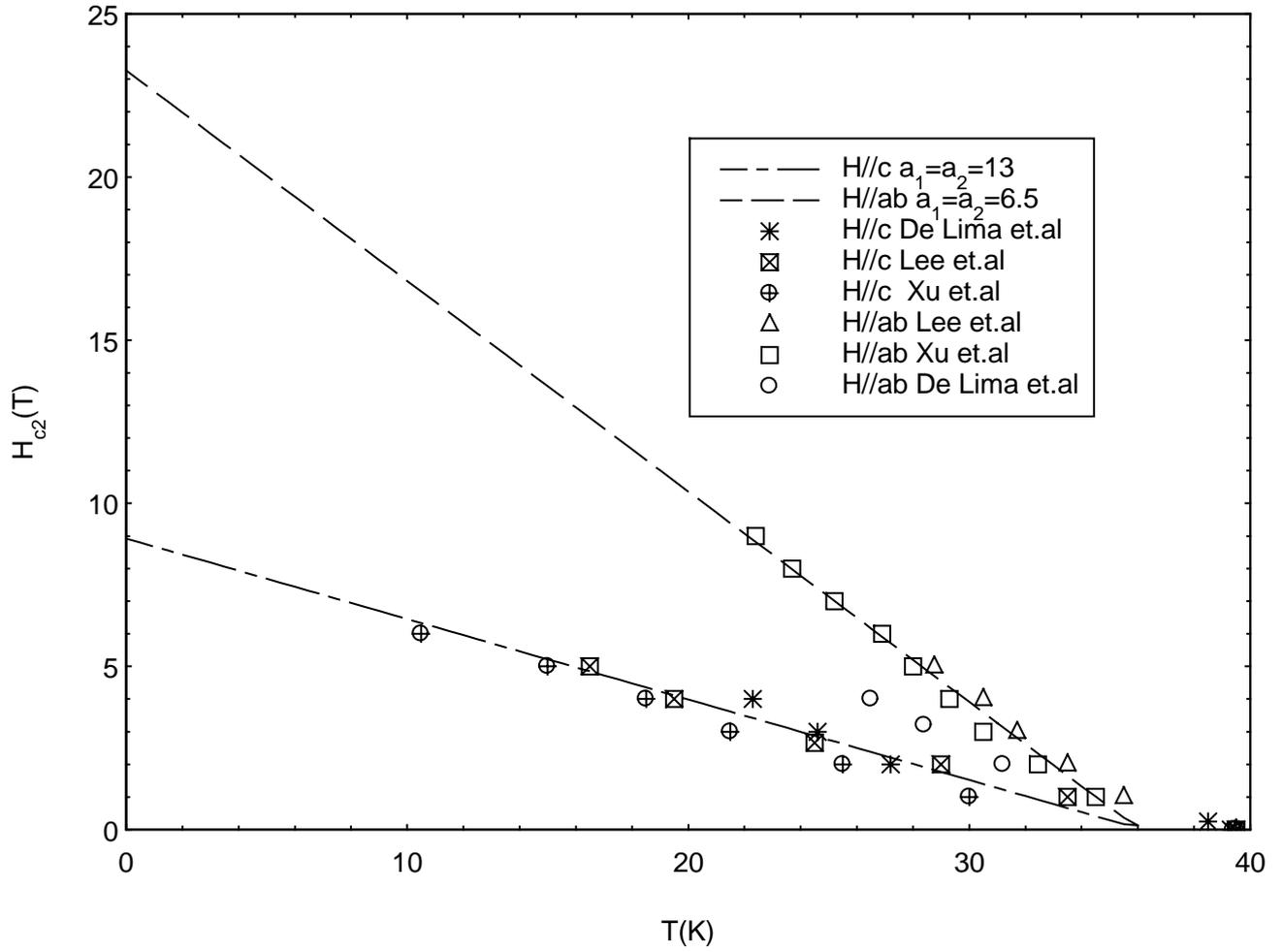

Figure(1). We fit Eq.(12) to the experimental data of upper critical field of single crystal MgB$_2$[21-23] and find the upper critical field in ab-plane($H_{c2}^{//ab}$), $H_{c2}^{//ab} = H_{c2}(6.5, 6.5, 0.7, 0.5)$, and in c-axis ($H_{c2}^{//c}$), $H_{c2}^{//c} = H_{c2}(13, 13, 0.7, 0.5)$.

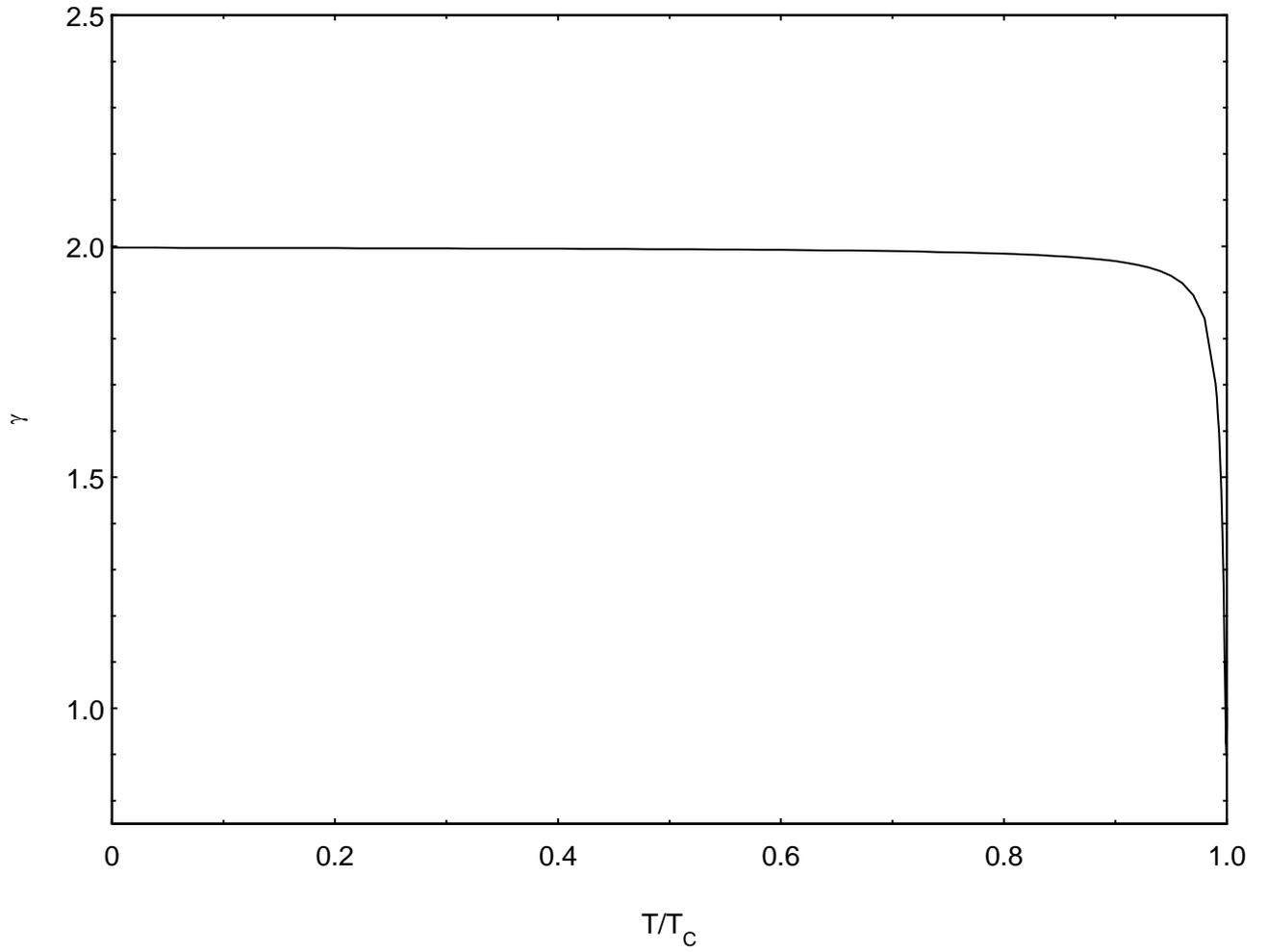

Figure(2). The ratio of upper critical field ( $\gamma = \dfrac{H_{c2}^{//ab}}{H_{c2}^{//c}}$ ) of the results of $H_{c2}$ in Figure.(1) versus temperature.